\documentclass[12pt,preprint]{aastex}
\usepackage{color}
\usepackage{ulem}

\shorttitle{CORONAL MAGNETIC HELICITY IN AR 10930}
\shortauthors{PARK et al.}
\begin{document}

\title{TIME EVOLUTION OF CORONAL MAGNETIC HELICITY IN THE FLARING ACTIVE REGION NOAA 10930}

\author{SUNG-HONG PARK,\altaffilmark{1} JONGCHUL CHAE,\altaffilmark{2} JU JING,\altaffilmark{1} CHANGYI TAN,\altaffilmark{3,4} AND HAIMIN WANG\altaffilmark{1}}

\altaffiltext{1}{Space Weather Research Laboratory, New Jersey Institute of Technology, 323 Martin Luther King Boulevard, 101 Tiernan Hall, Newark, NJ 07102, USA; sp295@njit.edu.}
\altaffiltext{2}{Astronomy Program and FPRD, Department of Physics and Astronomy, Seoul National University, Seoul 151-742, Republic of Korea.}
\altaffiltext{3}{IMSG Inc., Camp Springs, MD 20746, USA.}
\altaffiltext{4}{Center for Satellite Applications and Research, National Environmental Satellite, Data and Information Service, NOAA, Camp Springs, MD 20746, USA.}

\begin{abstract}
To study the three-dimensional (3D) magnetic field topology and its long-term evolution associated with the X3.4 flare of 2006 December 13, we investigate the coronal relative magnetic helicity in the flaring active region (AR) NOAA 10930 during the time period of December 8--14. The coronal helicity is calculated based on the 3D nonlinear force-free magnetic fields reconstructed by the weighted optimization method of Wiegelmann, and is compared with the amount of helicity injected through the photospheric surface of the AR. The helicity injection is determined from the magnetic helicity flux density proposed by Pariat et al. using $Solar$ $and$ $Heliospheric$ $Observatory$/Michelson Doppler Imager magnetograms. The major findings of this study are the following. (1) The time profile of the coronal helicity shows a good correlation with that of the helicity accumulation by injection through the surface. (2) The coronal helicity of the AR is estimated to be $-$4.3$\times$10$^{43}$ Mx$^2$ just before the X3.4 flare. (3) This flare is preceded not only by a large increase of negative helicity, $-$3.2$\times$10$^{43}$ Mx$^2$, in the corona over $\sim$1.5 days but also by noticeable injections of positive helicity though the photospheric surface around the flaring magnetic polarity inversion line during the time period of the channel structure development. We conjecture that the occurrence of the X3.4 flare is involved with the positive helicity injection into an existing system of negative helicity.
\end{abstract}


\section{INTRODUCTION}
The photospheric magnetic fields in the active region (AR) NOAA 10930 have been observed comprehensively by the Michelson Doppler Imager (MDI; Scherrer et al. 1995) on board the \textit{Solar and Heliospheric Observatory} (\textit{SOHO}) spacecraft and the Solar Optical Telescope (SOT; Tsuneta et al. 2008) on board the \textit{Hinode} satellite. In recent years, following the observations, considerable attention has been paid in the investigation of the structure of magnetic field lines and its evolution in AR 10930 related to the occurrence of the X3.4 flare on 2006 December 13. There were studies of sunspot rotation associated with the flare such as the remarkable counterclockwise rotation of the positive polarity sunspot (Yan et al. 2009), interaction between the fast rotating positive sunspot and ephemeral regions near the sunspot (Zhang et al. 2007), and nonpotential magnetic stress (Su et al. 2008). AR 10930 was also investigated for a change of magnetic field lines at the flaring site before and after the flare, e.g., in the azimuth angle (Kubo et al. 2007). Moreover, time variations of the magnetic helicity injection rate (Zhang et al. 2008; Magara \& Tsuneta 2008) and intermittency (Abramenko et al. 2008) were examined over a time span of several days around the time of the flare.

To resolve the limitations of using photospheric magnetic field data, some studies have been carried out on the X3.4 flare with the three-dimensional (3D) coronal magnetic fields derived from nonlinear force-free (NLFF) extrapolation methods. Jing et al. (2008) reported that magnetic shear around the flaring magnetic polarity inversion line decreased after the flare at coronal heights in the range of 8--70 Mm. By calculating the 3D electric current in AR 10930, Schrijver et al. (2008) showed that there are long fibrils of strong current slightly above the photosphere that almost completely disappear after the flare. Later on, Wang et al. (2008) found that the strong current-carrying fibrils are associated with the magnetic channel structure of AR 10930 and the flare occurred during the period in which the channels rapidly developed. In addition, the free energy of the NLFF fields was studied to understand the energy buildup, storage, and release processes in the corona during the flare. The free energy release of 2.4$\times$10$^{31}$ erg during the flare was measured by Guo et al. (2008), and Jing et al. (2010) found that a significant amount of free energy continuously built up in the 2 days prior to the flare.

Encouraged by interesting results of previous studies with NLFF fields, in this study, we investigate the variation of the coronal relative magnetic helicity in AR 10930 over a span of several days to determine its relationship with the flare.
Magnetic helicity is a measure of how much the magnetic field lines in a flux tube are twisted around the tube axis, how much the tube axis is kinked, and how much the flux tubes are interlinked with each other in a magnetic field system. It has been studied in order to understand the energy buildup process and trigger mechanism for flare occurrence. We anticipate that the coronal magnetic helicity study will bring a better understanding of the long-term evolution of the large-scale magnetic field geometry in the corona related to the X3.4 flare despite of a critical assessment (e.g., De Rosa et al. 2009) in NLFF extrapolation that existing NLFF extrapolation models are not able to accurately reproduce coronal fields and physical quantities of interest in the AR corona due to problematic issues such as the non-force-free nature of the photospheric magnetic field, limited field of view (FOV), and noise level of vector magnetograms, etc. Coronal helicity will also be compared with the helicity injection through the photospheric surface to check their relationship and consistency.

\section{CALCULATION OF MAGNETIC HELICITY}
The relative magnetic helicity, $H_{r}$, derived by Finn \& Antonsen(1985) is used to calculate a topologically meaningful and gauge-invariant measure of helicity inside a volume, $V$:
\begin{equation}
H_{r}=\int_{V} (\textit{\textbf{A}}+\textit{\textbf{A}}_{p}) \cdot (\textit{\textbf{B}}-\textit{\textbf{P}})\,dV,
\end{equation}
where $\textit{\textbf{P}}$ is the potential field having the same normal component as the magnetic field, $\textit{\textbf{B}}$, on the boundary surface enclosing $V$. $\textit{\textbf{A}}$ and $\textit{\textbf{A}}$$_{p}$ are the vector potentials for $\textit{\textbf{B}}$ and $\textit{\textbf{P}}$, respectively. $H_{r}$ represents the amount of helicity subtracted from the corresponding potential field $\textit{\textbf{P}}$. In our calculation of $H_{r}$ in a coronal volume of AR 10930, we adopt the code of Fan (2009) for the determination of the specific vector potentials, $\textit{\textbf{A}}$ and $\textit{\textbf{A}}$$_{p}$, proposed by DeVore (2000) in treating the photosphere as an infinite plane ($z$ = 0) in a Cartesian coordinate system. The unsigned magnetic flux, $\Phi$, through the photospheric surface, $S$, of AR 10930 is defined by
\begin{equation}
\Phi = \int_{S} \left|B_z\right| \, dS,
\end{equation}
where $B_{z}$ is the $z$-component of magnetic field and the integration is over the entire photospheric area ($z$ = 0) of the computational domain of 3D NLFF field data.
Note that outside of the computational domain of AR 10930, the magnetic field is assumed to be negligible, even though, on average, $\sim$30\% of $\Phi$ passed through the domain of the actual 3D NLFF fields. Our helicity calculation, therefore, gives an approximate value of $H_{r}$ in a coronal volume above the photospheric surface of AR 10930.

Throughout this paper, by magnetic helicity we mean the relative magnetic helicity. We estimate the rate of magnetic helicity injection, $\dot{H}_{r}$, into the coronal volume through $S$ of AR 10930 using the method developed by Chae (2007):
\begin{equation}
\dot{H}_{r}=\int_{S} G_{\theta} \,dS,
\end{equation}
where $G_{\theta}$ is the helicity flux density proposed by Pariat et al. (2005), and can be obtained with the normal component of magnetic field and the apparent horizontal velocity, $\textit{\textbf{u}}$, of the photospheric field line footpoints. We determine $\textit{\textbf{u}}$ by applying the normal component of the magnetic induction equation and the differential affine velocity estimator (DAVE) method developed by Schuck (2006). Please refer to the procedure described in Chae (2007) for the details of the $\dot{H}_{r}$ calculation. After $\dot{H}_{r}$ is determined as a function of time, we integrate it with respect to time to determine the amount of helicity accumulation, $\Delta H_{r}$:
\begin{equation}
\Delta H_{r} = \int_{t_0} ^{t} \dot{H}_{r}\,dt,
\end{equation}
where $t_0$ and $t$ are the start and end time of the data set under investigation, respectively.

\section{DATA PREPARATION AND PROCESSING}
For the calculation of $H_{r}$ in a 3D coronal volume, the three components of the magnetic field in that coronal volume need to be obtained. Therefore, we follow the same method described in Jing et al. (2010) in deriving the coronal NLFF fields in AR 10930 from the Stokes profiles taken by the \textit{Hinode}/SOT Spectro-Polarimeter (SP). We first derived the high-resolution vector magnetic fields in the photosphere from the Stokes profiles using an Unno--Rachkovsky inversion based on the Miler--Eddington atmosphere (e.g., Lites \& Skumanich 1990; Klimchuk et al. 1992). In addition, the removal of the 180$^{\circ}$ ambiguity in the transverse magnetic fields was accomplished using the minimum energy algorithm (Metcalf et al. 2006), and the photospheric vector magnetograms were projected onto the tangent plane at the heliographic location of the center of the magnetograms. To reduce the inaccuracy of NLFF field extrapolation, it is important to derive suitable boundary fields for the NLFF field modeling from the photospheric magnetograms. Therefore, using a preprocessing method developed by Wiegelmann et al. (2006), we minimized the effect of the Lorentz force acting in the photosphere and prepared the NLFF boundary fields to be in the condition of the low plasma-$\beta$ force-free chromosphere. We then used the weighted optimization method (Wiegelmann 2004) to extrapolate the NLFF coronal fields from the photospheric magnetograms. This method has been well recognized as an outstanding performance algorithm by some model tests of NLFF fields (e.g., Schrijver et al. 2006; Metcalf et al. 2008).

AR 10930 appeared on the east limb of the solar disk on 2006 December 6, and was successfully and continuously observed during the time interval of its entire disk passage by \textit{Hinode}/SOT and \textit{SOHO}/MDI. In this study, we determine $H_{r}$ in AR 10930 during the time span of 2006 December 8, 21:20 UT through 2006 December 14, 5:00 UT using 27 \textit{Hinode}/SOT-SP vector magnetograms as the boundary fields for NLFF field extrapolation. The computational dimensions of the 3D NLFF field data were considered as 240$\times$132$\times$180 pixel$^{3}$ corresponding to 288$\times$158$\times$216 Mm$^{3}$. To check the influence of preprocessing on the magnetogram data, we calculated $L_{1}$ and $L_{2}$ of the original data and those of the preprocessed data which Wiegelmann et al. (2006) proposed to investigate in order to determine how well a photospheric magnetic field agrees with Aly's criteria: $L_{1}$ and $L_{2}$ are related to the force-balance condition and the torque-free condition, respectively. Refer to Wiegelmann et al. (2006) for the details of the preprocessing method and the definitions of the $L_{1}$ and $L_{2}$. As shown in Table 1, the preprocessed data satisfy the Aly criteria much better than the original data. It has been reported that this preprocessing procedure significantly improves the boundary fields toward a force-free condition (e.g., Wiegelmann et al. 2006, 2008). Recently, Jing et al. (2010) also showed the capability of the preprocessing method by comparing the unpreprocessed/preprocessed photospheric line of sight (LOS) magnetogram of AR 10930 with the co-aligned chromospheric LOS magnetogram. To evaluate the performance of the NLFF extrapolation, we also calculated the current-weighted sine metric (CWsin) and the $\langle$$|$$f_{i}$$|$$\rangle$ metric proposed by Wheatland et al. (2000) for each extrapolated field. CWsin and $\langle$$|$$f_{i}$$|$$\rangle$ measure the degree of convergence to a force-free and divergence-free field, respectively. For the 27 NLFF fields under investigation, the average CWsin was estimated as $\sim$0.39 and the average $\langle$$|$$f_{i}$$|$$\rangle$ as $\sim$0.0014 indicating that residual forces and divergences exist in the NLFF fields.

In addition, the error estimation of $H_{r}$ is carried out with a Monte Carlo method by only taking into account the sensitivity of the SP measurement as follows (e.g., see Guo et al. 2008): first, we add three sets of artificial noises to $B_{x}$, $B_{y}$, and $B_{z}$ of the original SP vector magnetogram at 20:30 UT on 2006 December 12. Each noise set consists of pseudorandom numbers in normal distribution with the standard deviation of 5 G for $B_{z}$ and 50 G for $B_{x}$ and $B_{y}$. Note that these values of 5 G and 50 G are estimated as the maximum values of the SP sensitivity in the LOS direction and the transverse direction, respectively (Tsuneta et al. 2008). Then, we extrapolate the 3D NLFF fields from the noise-imposed vector magnetogram following the procedure described in the above paragraph, calculate $H_{r}$, and repeat the same process 10 times. Finally, we consider the standard deviation of 10 sets of $H_{r}$ to be the uncertainty of the $H_{r}$ calculation. The uncertainty was found to be 8$\times$10$^{41}$ Mx$^2$ corresponding to 2$\%$--4$\%$ of $|H_{r}|$ during the measurement period.

In order to calculate $\dot{H}_{r}$, we used the data set consisting of 63 full-disk MDI magnetograms at the 96 minute cadence in the time span of 2006 December 8, 20:51 UT through 2006 December 13, 16:03 UT. Note that the MDI magnetograms in the data set show the Zeeman saturation in the central part of the negative sunspot umbral region, which means that our calculation of $\dot{H}_{r}$ might be underestimated. The window function of DAVE used in the $\dot{H}_{r}$ calculation is the top-hat profile which puts the same weight of unity to every pixel inside the window (e.g., Schuck 2006) and the window size is selected to be 10 arcsec. We also applied DAVE to two MDI images with the spatial derivatives calculated from the average of the two images (e.g., Welsch et al. 2007; Chae 2007; Chae \& Sakurai 2008). The uncertainty of $\dot{H}_{r}$ corresponding to the measurement uncertainty ($\sim$20 G) of MDI magnetograms was also estimated using the same Monte Carlo method used in the error estimation of $H_{r}$. It is found that the uncertainty of $\dot{H}_{r}$ is 8.4$\times$10$^{39}$ Mx$^2$ hr$^{-1}$ which is equivalent to $\sim$3\% of the average $\dot{H}_{r}$ during the measurement time. The uncertainty therefore does not significantly affect our study of $\dot{H}_{r}$ and $\Delta H_{r}$.

\section{RESULT AND DISCUSSION}
Our main objective in this study is to examine how well $H_{r}$ and $\Delta H_{r}$ are correlated with each other and whether our $H_{r}$ calculation using the NLFF coronal fields is verified through a comparison of the $H_{r}$ derived from the \textit{Hinode}/SOT-SP data with the $\Delta H_{r}$ derived from the \textit{SOHO}/MDI data. In Figure 1, therefore, we plot the temporal variations of $H_{r}$ (black solid line) and $\Delta H_{r}$ (gray solid line). The estimated error in $H_{r}$ is marked with error bars. The initial value of $\Delta H_{r}$ is set the same as that of $H_{r}$. $|H_{r}|$, the absolute value of $H_{r}$, is also shown by a dotted line for convenience.
We also investigate the day-to-day variations of $H_{r}$ in AR 10930 for a better understanding of the pre-flare conditions and a trigger mechanism of the X3.4 flare. For this, $H_{r}$ (black solid line) is plotted with the total unsigned magnetic flux (dashed line) and the \textit{GOES} soft X-ray light curve (dotted line) in Figure 2.
Note that Lim et al. (2007) have done a similar study in which they compared the coronal helicity in AR 10696 with the helicity injection through the photosphere. In their study, the coronal helicity was estimated as a probable range using a linear force-free (LFF) assumption with a force-free constant that gives the best fit with each of the individual coronal loops, even though the real coronal field is not LFF. The photospheric helicity injection was calculated by inferring the velocity of the apparent horizontal motion of the field lines determined by the technique of local correlation tracking (LCT), as originally proposed by Chae (2001), instead of using $\textit{\textbf{u}}$ determined by the DAVE technique. They found that the temporal variation of the coronal helicity is similar to that of the photospheric helicity injection with a discrepancy of $\sim$15\%.

During the first day of the helicity measurement, $H_{r}$ showed little change from its initial value, $-$2.8$\times$10$^{43}$ Mx$^2$, though there were small fluctuations in the range of 2$\%$--15$\%$. Then, $|H_{r}|$ decreased by 28$\%$ from 2.9$\times$10$^{43}$ Mx$^2$ to 2.1$\times$10$^{43}$ Mx$^2$ for 14 hr from December 10. Note that the decrease of $|H_{r}|$ could be due to (1) a pre-existing negative helicity being expelled from the volume of the NLFF field extrapolation, e.g., via coronal mass ejections (CMEs), and/or (2) a new magnetic flux with positive helicity being injected from outside into the volume or a positive helicity being produced by the shearing motions of pre-existing field lines. We found that there are three time periods (I, II$_{b}$, and III) over which $|H_{r}|$ decreases consistently for more than nine hours, and they are shown as shaded areas in Figure 1. Between periods I and III, there was a consistently large increase of negative helicity, $-$3.2$\times$10$^{43}$ Mx$^2$, in the corona over $\sim$1.5 days (marked as period II$_{a}$ in Figure 1). After period III, a negative helicity kept on increasing for $\sim$1 day with flux increase. The detailed information of the characteristic periods is shown in Table 2.

We compare the overall pattern of the temporal evolution of the $H_{r}$ calculated using the NLFF fields with that of the $\Delta H_{r}$ measured using the MDI magnetograms. In general, the time profile of $H_{r}$ matches well that of $\Delta H_{r}$. Moreover, in both cases, the absolute amount of negative helicity accumulation during the entire measurement period of December 9--14 was similar (2.1$\times$10$^{43}$ Mx$^2$ and 1.7$\times$10$^{43}$ Mx$^2$, respectively). This gives us confidence that the NLFF extrapolation and the $H_{r}$ calculation are reasonably well established. However, some detailed patterns of helicity evolution show a difference between $H_{r}$ and $\Delta H_{r}$. For example, the temporal variation of $H_{r}$ shows a rapid and large increase of negative helicity with flux increase at the time period of the fast rotational speed in the southern positive sunspot measured by Min \& Chae (2009) and Yan et al. (2009). In addition, $|H_{r}|$ represents decreasing phases such as periods I, II$_{b}$, and III, while $|\Delta H_{r}|$ increases monotonically during the entire period. Note that $H_{r}$ should not necessarily be exactly the same as $\Delta H_{r}$: e.g., the ejection of magnetic helicity via the launch of a CME would not be detected in $\Delta H_{r}$ while it would be reflected in $H_{r}$.

What could cause the three periods of remarkable $|H_{r}|$ decrease? To investigate this, we first checked a possibility associated with the negative helicity ejection via CMEs that originated from AR 10930. The \textit{SOHO}/Large Angle and Spectrometric Coronagraph (LASCO; Yashiro et al. 2004) CME catalog was used to search for all the CMEs that occurred during the three periods. We then identified only the CMEs inferred to be produced in AR 10930 with the following criterion: the position angle of a CME should be within $\pm$5$^{\circ}$ from that of AR 10930 on the solar disk at the first appearance time of the CME in the LASCO/C2 FOV. Note that there were no other ARs except for AR 10930 on the front side of the solar disk during the periods. We found two CMEs: one in period II$_{b}$ and the other in period III. Their initial appearances in the LASCO/C2 FOV were at 09:36 UT on December 11 and at 20:28 UT on December 12, respectively, which are marked with the vertical dashed lines in Figure 2. Although the uncertainty of our $H_{r}$ calculation is estimated to be 8$\times$10$^{41}$ Mx$^2$, we found that the decrease in $|H_{r}|$ is 2.4$\times$10$^{42}$ Mx$^2$ between 08:31 UT and 11:48 UT on December 11 and 1.9$\times$10$^{42}$ Mx$^2$ between 18:12 UT and 21:01 UT on December 12 covering the time of the occurrence of the first CME and that of the second CME, respectively. These values agree with the helicity content of a typical CME, 2$\times$10$^{42}$ Mx$^2$, estimated by DeVore (2000). Our finding of the CME-related change of $|H_{r}|$ is similar to the earlier finding by Lim et al. (2007) in which they found a helicity decrease of $\sim$4.1$\times$10$^{42}$ Mx$^2$ after the occurrence of two CMEs.

We also investigated the feasibility of positive helicity injection through the photospheric surface of AR 10930 into the corona.
Note that Zhang et al. (2008) have calculated $\dot{H}_{r}$ in AR 10930 using the LCT method (Chae 2001). They found that the sign of $\dot{H}_{r}$ changes from negative to positive and then from positive to negative during the period (01:30 UT--04:30 UT) of the flare, while $\dot{H}_{r}$ is predominantly negative during 2006 December 8--14. Integrating the positive (negative) $G_{\theta}$ over the photospheric surface of AR 10930, we determined $\dot{H}_{r}^{+}$ ($\dot{H}_{r}^{-}$), i.e., the injection rate of positive (negative) helicity. Figure 3 shows the time variations of $\dot{H}_{r}^{+}$ (diamonds), $\dot{H}_{r}^{-}$ (crosses), and $\dot{H}_{r}$ (solid line) during the $\Delta H_{r}$ measurement period. The characteristic periods are marked in the same way as in Figure 1, and the peak time of the X3.4 flare is shown as the vertical dotted line. We found that a remarkable accumulation of positive helicity into the corona is established over the entire period with an average injection rate of 2.8$\times$10$^{41}$ Mx$^2$ hr$^{-1}$, even though most of the time $\dot{H}_{r}^{-}$ is dominant with an average injection rate of -4.4$\times$10$^{41}$ Mx$^2$ hr$^{-1}$. Especially during the span of December 11, 12:51 UT (middle of period II$_{b}$) through December 12, 04:48 UT (start of period III), the average of $\dot{H}_{r}^{+}$ showed a large value of 4.5$\times$10$^{41}$ Mx$^2$ hr$^{-1}$, and $\dot{H}_{r}^{+}$ was sometimes larger than $\dot{H}_{r}^{-}$. Additionally, we examined the $G_{\theta}$ maps at several times (marked by vertical solid lines in Figure 3) to find out how the positive $G_{\theta}$ is distributed and developed on the AR. Figure 4 shows the maps of the normal component of the magnetic field, $B_{n}$ (left panels), and $G_{\theta}$ (right panels). Assuming that the magnetic field on the solar photosphere is normal to the solar surface, $B_{n}$ was approximately determined from the MDI LOS magnetograms. We found that there are noticeable injections of positive helicity around the flaring magnetic polarity inversion line (see the three $G_{\theta}$ maps in Figure 4: 2006-12-11 12:51 UT, 2006-12-12 04:48 UT, and 2006-12-12 23:59 UT). In addition, the examination of the other $G_{\theta}$ maps during the period of December 11, 12:00 UT through December 13, 16:00 UT revealed that positive helicity is consistently injected through the polarity inversion line. The location and time span of the positive helicity injection are similar to those of the magnetic channel structure development observed by Wang et al. (2008). Note that a simulation by R{\'e}gnier (2009) shows that newly injected current from the photosphere can sensitively affect the coronal magnetic helicity in existing force-free bipolar fields: i.e., $H_{r}$ is increased by 2 orders of magnitude when the current strength is increased by a factor of 2. We therefore speculate that periods II$_{b}$ and III are associated with the helicity ejection via the two CMEs and/or the supply of positive helicity from the photosphere into the corona.

Related to the occurrence of the X3.4 flare, we found two interesting patterns of the long-term $H_{r}$ evolution. First, there was a significant increase of negative $H_{r}$ for period II$_{a}$ of $\sim$1.5 days associated with the flare energy buildup. This pattern of increasing helicity prior to the flare is in agreement with that shown in the study of Park et al. (2008, 2010). After the middle of period II$_{a}$, a large amount of helicity of the opposite (positive) sign started to be injected through the photospheric surface around the flaring magnetic polarity inversion line during the time span (including periods II$_{b}$ and III) of the channel structure development observed by Wang et al. (2008). The X3.4 flare was preceded by the two characteristic patterns of $H_{r}$. These two patterns have been already reported by previous studies of major flares related with helicity injection through photospheric surfaces of ARs (Park et al. 2008, 2010; Chandra et al. 2010). Note that our finding of the long-term injection of positive helicity $\sim$2.5 days before the flare is different from the abrupt injection of positive helicity around the start of the flare found by Zhang et al. (2008). We conjecture that the occurrence of the X3.4 flare is involved with the emergence of a positive helicity system into an existing negative helicity system which may cause a reconnection between the two helicity systems. This idea is not only supported by numerical simulation (Kusano et al. 2003b) in which magnetic reconnection quickly grows in the site of helicity annihilation with different signs but also by observational reports for the opposite sign of helicity injection through the photosphere surface of ARs before flares (Kusano et al. 2003a; Yokoyama et al. 2003; Wang et al. 2004).

In conclusion, after analyzing $H_{r}$ in the coronal volume of AR 10930 using the NLFF fields, we found that there are two characteristic phases of day-to-day variation of helicity related to the X3.4 flare: significant helicity accumulation (period II$_{a}$) followed by opposite sign helicity injection (periods II$_{b}$ and III). $H_{r}$ and $\Delta H_{r}$ show a roughly similar variation during the entire measurement period. Further studies are needed to check whether the two characteristic patterns are shown in other major flaring ARs and to investigate the short-term variation of helicity in a flaring region related to a triggering mechanism. The \textit{Solar Dynamic Observatory} (\textit{SDO}) was recently launched and we expect to study the 3D coronal helicity using full-disk photospheric vector magnetograms with high spatial and temporal resolution taken by the Helioseismic and Magnetic Imager (HMI) on board the \textit{SDO}.

\acknowledgments
We are grateful to the referee for helpful and constructive comments. The authors thank Dr. Yuhong Fan for sharing the code to determine the 3D vector potential, and Dr. Thomas Wiegelmann for providing the weighted optimization and preprocessing codes for NLFF field extrapolation. \textit{SOHO} is a project of international cooperation between ESA and NASA. \textit{Hinode} is a Japanese mission developed and lunched by ISAS/JAXA, collaborating with NAOJ as a domestic partner, NASA and STFC (UK) as international partners. It is operated by these agencies in co-operation with ESA and NSC (Norway). This work was supported by the National Research Foundation of Korea (KRF-2008-220-C00022). J.J. was supported by NSF under grants ATM 09-36665 and ATM 07-16950. C.T. was supported by DLR-grant 50 OC 0501 and the Office of Sponsored Program, NJIT. S.-H.P. and H.W. were supported by NSF grant AGS-0745744 and NASA grant NNX08BA22G.

\clearpage


\begin{deluxetable}{lcc}
\tabletypesize{\scriptsize}
\tablewidth{0pt}
\tablecaption{Comparison of the Average $L$-values for the Original and Preprocessed \textit{Hinode}/SOT-SP Vector Magnetograms \label{1}}
\tablehead{
\colhead{} & \colhead{Original Data} & \colhead{Preprocessed Data}}
\startdata
$L_{1}$\tablenotemark{a} (G$^{4}$)          & 1.12$\times$10$^{19}$ & 9.56$\times$10$^{12}$ \\
$L_{2}$\tablenotemark{b} (G$^{4}$ Mm$^{2}$)   & 2.02$\times$10$^{23}$ & 1.08$\times$10$^{19}$ \\[0.5ex]
\enddata
\tablenotetext{a}{$L_{1}$=$\left[\left(\Sigma B_{x}B_{z}\right)^{2}+\left(\Sigma B_{y}B_{z}\right)^2+\left(\Sigma B_{z}^{2}-B_{x}^{2}-B_{y}^{2}\right)^2\right]$}
\tablenotetext{b}{$L_{2}$=$\left[\left(\Sigma x(B_{z}^{2}-B_{x}^{2}-B_{y}^{2})\right)^{2}+\left(\Sigma y(B_{z}^{2}-B_{x}^{2}-B_{y}^{2})\right)^{2}+\left(\Sigma yB_{x}B_{z}-xB_{y}B_{z}\right)^2\right]$}
\end{deluxetable}

\begin{deluxetable}{lccccc}
\tabletypesize{\scriptsize}
\tablewidth{0pt}
\tablecaption{Characteristic Periods of the Temporal Variation of the Coronal Magnetic Helicity \label{2}}
\tablehead{
\colhead{Periods} & \colhead{Duration} & \colhead{Initial/Final $|H_{r}|$} & \colhead{$|H_{r}|$ Change} & \colhead{Initial/Final Flux} & \colhead{Flux Change}\\
\colhead{} & \colhead{(hr)} & \colhead{(10$^{43}$ Mx$^{2}$)} & \colhead{(\%)} & \colhead{(10$^{22}$Mx)} & \colhead{(\%)}}
\startdata
I & 13.8 & 2.9 / 2.1  & -28 & 5.2 / 5.0 & -4 \\
II$_{a}$ & 40.1 & 2.1 / 5.3 & 152 & 5.0 / 5.5 & 10 \\
II$_{b}$ & 9.0 & 4.8 / 4.0 & -17 & 5.6 / 5.3 & -5 \\
III & 16.6 & 5.3 / 4.3 & -19 & 5.5 / 5.6 & 2 \\[0.5ex]
\enddata
\end{deluxetable}


\clearpage


\begin{figure}
\begin{center}
\includegraphics[scale=0.9]{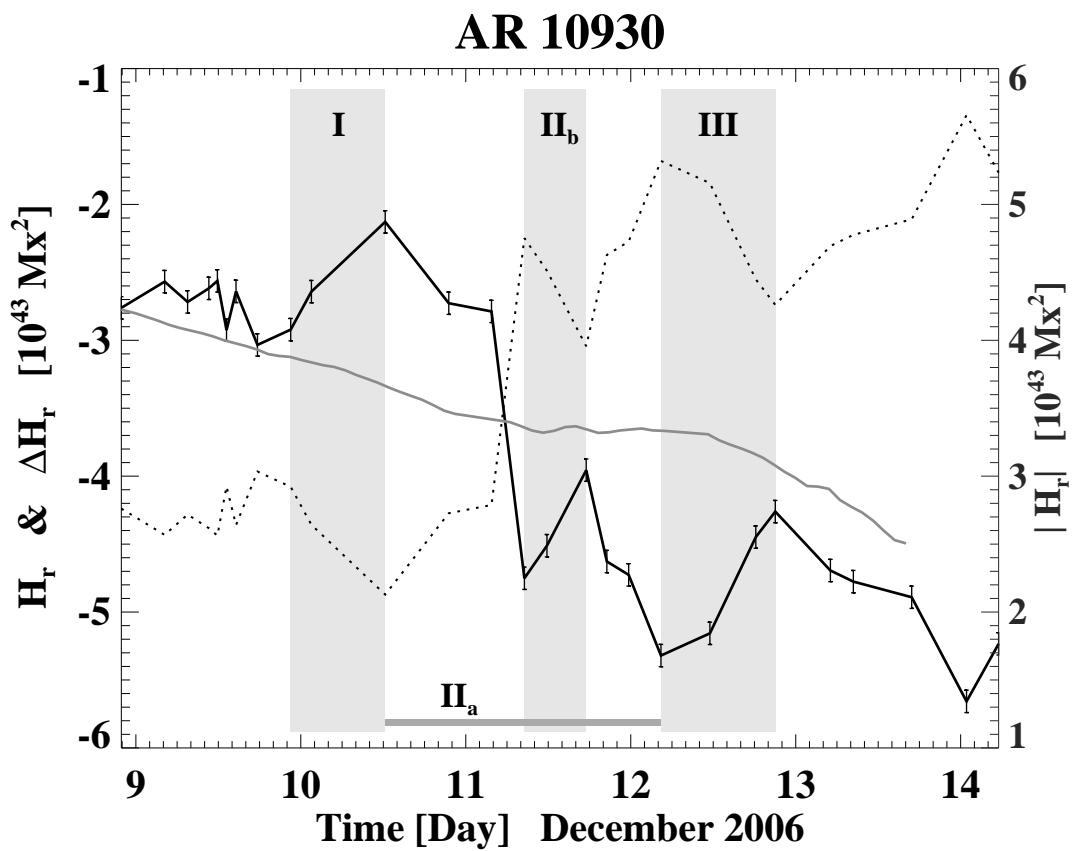}
\caption{Time variations of the coronal relative magnetic helicity $H_{r}$ (black solid line with error bars) and the helicity accumulation $\Delta H_{r}$ (gray solid line). The absolute value of $H_{r}$ decreases for more than 9 hr in the periods marked as I, II$_{b}$, and III while it shows a significant increase of 3.2$\times$10$^{43}$ Mx$^2$ during the period of II$_{a}$. In general, the time profile of $H_{r}$ shows a good correlation with that of $\Delta H_{r}$ during the entire measurement period.
\label{fig1}}
\end{center}
\end{figure}

\begin{figure}
\begin{center}
\includegraphics[scale=0.85]{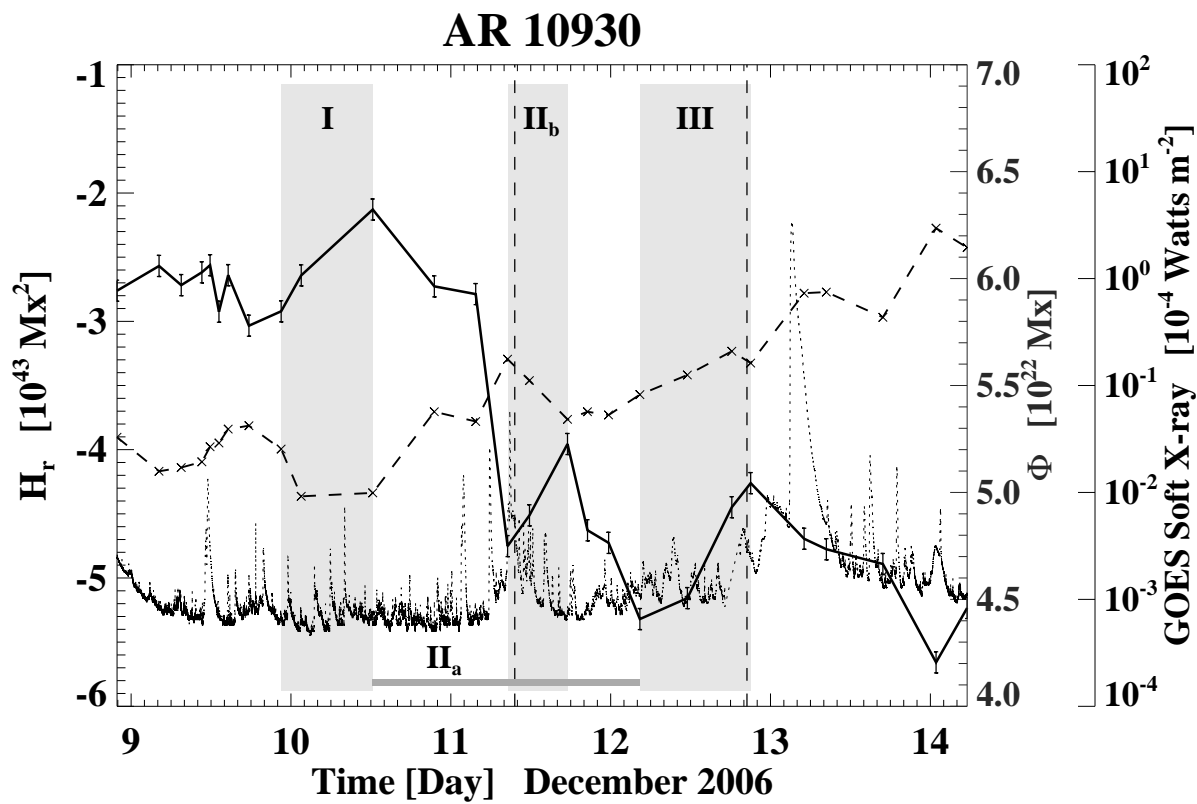}
\caption{Coronal relative magnetic helicity $H_{r}$ (black solid line with error bars) and the unsigned magnetic flux $\Phi$ (dashed line) of AR 10930 plotted with the \textit{GOES} soft X-ray flux (dotted line) during the time period of December 8, 21:20 UT through December 14, 5:00 UT. The X3.4 flare occurred in AR 10930 and peaked at 2:40 UT on 2006 December 13. During the periods of II$_{b}$ and III, there were two CMEs inferred to have originated from AR 10930, and their first appearance times in the LASCO/C2 FOV are marked with the black vertical dashed lines. The characteristic periods of I, II$_{a}$, II$_{b}$, and III are marked in the same way as in Figure 1.
\label{fig2}}
\end{center}
\end{figure}

\begin{figure}
\begin{center}
\includegraphics[scale=0.9]{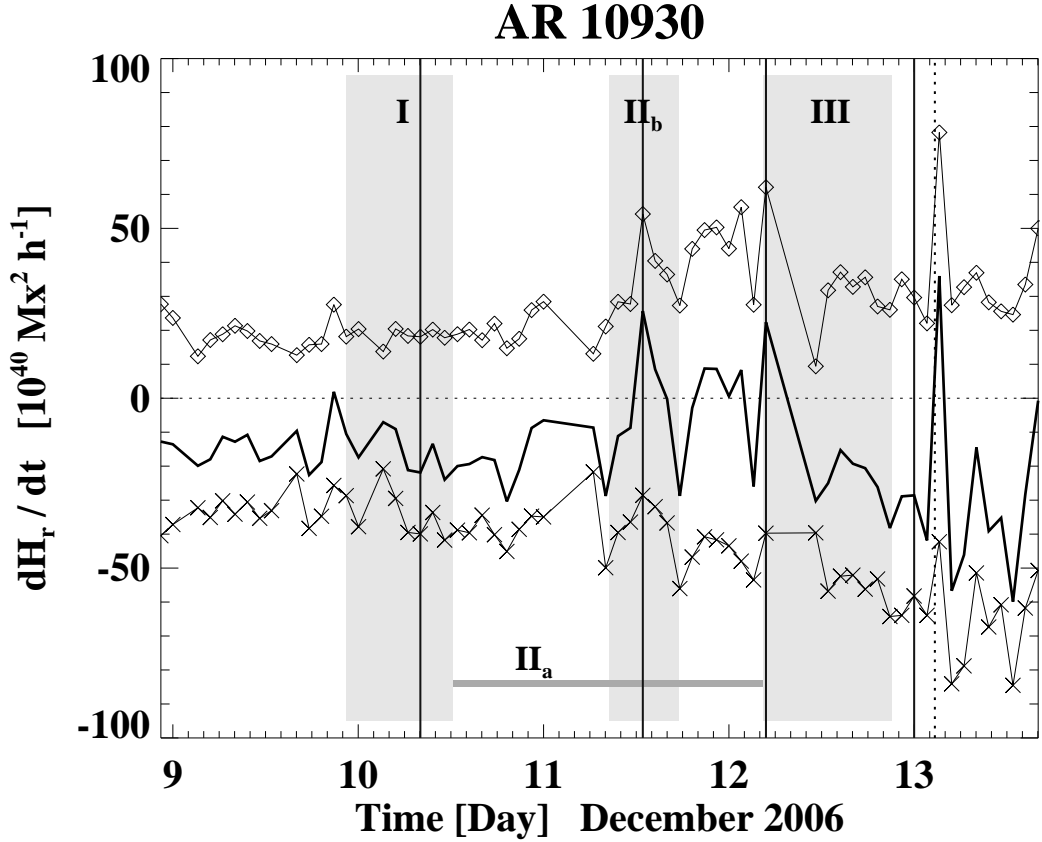}
\caption{Injection rates of positive helicity (diamonds), negative helicity (crosses), and total helicity (solid line) during the time span of December 8, 20:51 UT to December 13, 16:03 UT. The characteristic periods of I, II$_{a}$, II$_{b}$, and III are marked in the same way as in Figure 1, and the peak time of the X3.4 flare is shown by the vertical dotted line. The vertical solid lines indicate the times of the investigation of the helicity flux density maps in Figure 3.
\label{fig3}}
\end{center}
\end{figure}

\begin{figure}
\begin{center}
\includegraphics[scale=0.8]{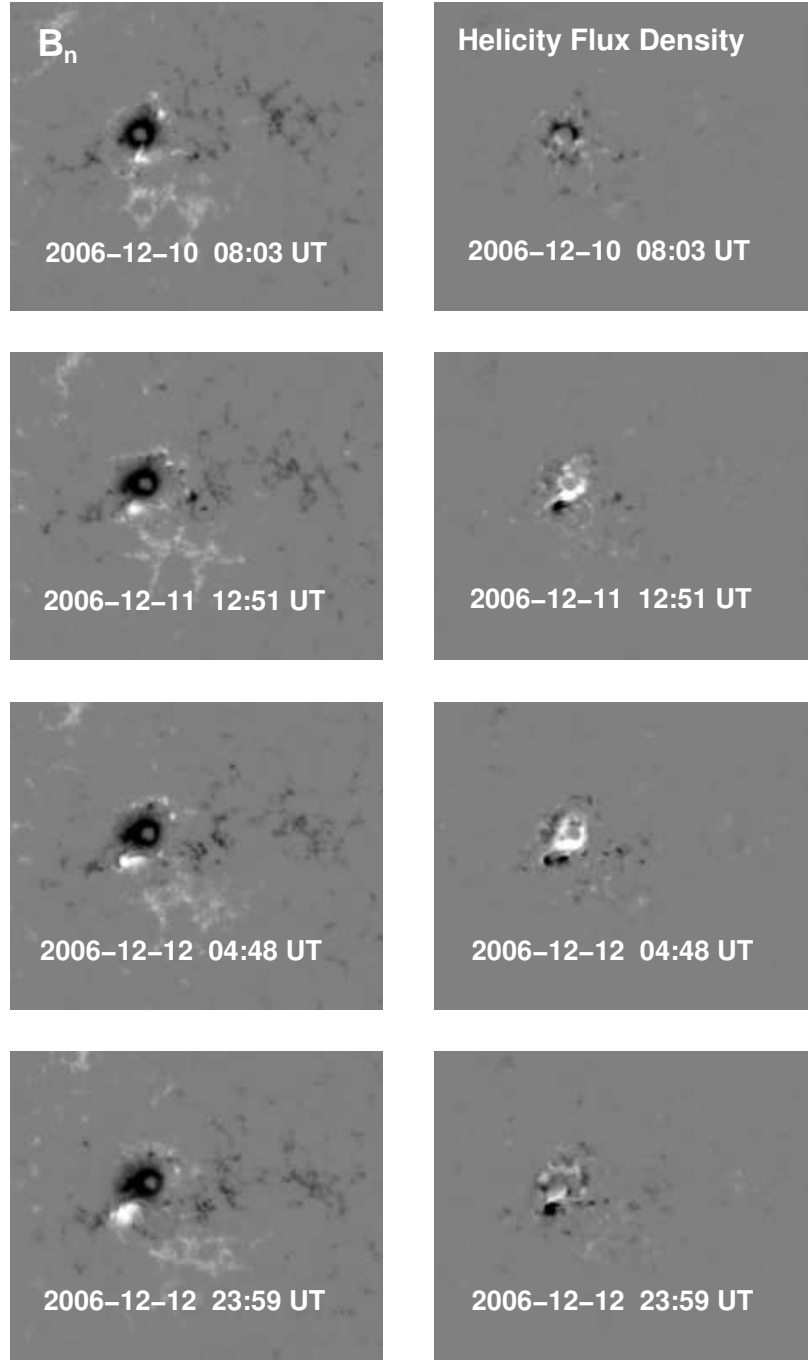}
\caption{Temporal evolution of the photospheric magnetic field and the helicity injection rate in AR 10930. Left panels: the normal component of the magnetic field $B_{n}$ derived from the MDI LOS magnetograms. Right panels: helicity flux density $G_{\theta}$. Note that the median of $|G_{\theta}|$ is $\sim$2$\times$10$^{3}$ G$^2$ km s$^{-1}$ Mm, and we set the saturation level of $|G_{\theta}|$ as 2.5$\times$10$^{6}$ G$^2$ km s$^{-1}$ Mm for purpose of display visibility. After the middle of December 11, a large amount of positive helicity started to be injected around the flaring magnetic polarity inversion line.
\label{fig4}}
\end{center}
\end{figure}


\end{document}